\documentclass{INTERSPEECH2023}
\usepackage{mathtools}
\usepackage{multirow}
\usepackage{xcolor}
\usepackage{url}

\title{MParrotTTS: Multilingual Multi-speaker Text to Speech Synthesis in Low Resource Setting}
\name{Neil Shah$^{1,2}$, Vishal Tambrahalli$^1$, Saiteja Kosgi$^1$, Niranjan Pedanekar$^2$, Vineet Gandhi$^1$}
\address{
  $^1$Kohli Centre on Intelligent Systems, IIIT Hyderabad, India\\
  $^2$TCS Research, Pune, India}
\email {\{neilkumar.shah, vishal.tambrahalli, saiteja.k\}@research.iiit.ac.in, n.pedanekar@tcs.com, vgandhi@iiit.ac.in}

\newcommand{\ourmodel}{MParrotTTS}
\newcommand{\khz}{kHz}
\newcommand{\hmb}[1]{\mathbf{h}_{#1}}
\newcommand{\enc}[1]{\mathbf{E}_{#1}}
\newcommand{\nar}{NAR}

\newcommand{\baselinefs}{FastSpeech2}
\newcommand{\baselineml}{Meta-TTS}
\newcommand{\etal}{\textit{et al.}}
\begin{document}

\maketitle
 
\begin{abstract}
We present \ourmodel, a unified multilingual, multi-speaker text-to-speech (TTS) synthesis model that can produce high-quality speech. Benefiting from a modularized training paradigm exploiting self-supervised speech representations, \ourmodel~adapts to a new language with minimal supervised data and generalizes to languages not seen while training the self-supervised backbone. Moreover, without training on any bilingual or parallel examples, \ourmodel~can transfer voices across languages while preserving the speaker-specific characteristics, e.g., synthesizing fluent Hindi speech using a French speaker’s voice and accent. We present extensive results on six languages in terms of speech naturalness and speaker similarity in parallel and cross-lingual synthesis. The proposed model outperforms the state-of-the-art multilingual TTS models and baselines, using only a small fraction of supervised training data. Speech samples from our model can
be found at \textcolor{blue}{\url{https://paper2438.github.io/tts/}} 

%Benefiting from self supervised speech representation, our model is able to generate speech using minimal amount of paired text-to-speech data. 

\end{abstract}

\section{Introduction}
\begin{minipage}{0.45\textwidth}
``With languages, you are at home anywhere.''
\end{minipage}
\\[3pt]
\rightline{{\rm -- Edward De Waal}} \\

Our work focuses on a unified Text-to-Speech (TTS) system that can communicate using various languages (multilingual) and different voices (multispeaker), facilitating cross-language synthesis, i.e., allowing multiple languages to be spoken in each of the speaker's voices. We aim to accomplish this objective by utilizing limited supervised data, enabling adaptation to languages with limited resources. The most straightforward way to obtain such a TTS (text-to-speech) model is to train it with a multilingual speech dataset for every speaker involved. However, even with rich-resource languages (RRLs), such data is not available, as it is hard to find speakers with native proficiency in several languages. Training multilingual TTS faces various other challenges, such as the need for distinct input representations or pronunciations and uneven distribution of training data across languages.

\begin{figure}[t]
  \centering
  \includegraphics[width=\linewidth]{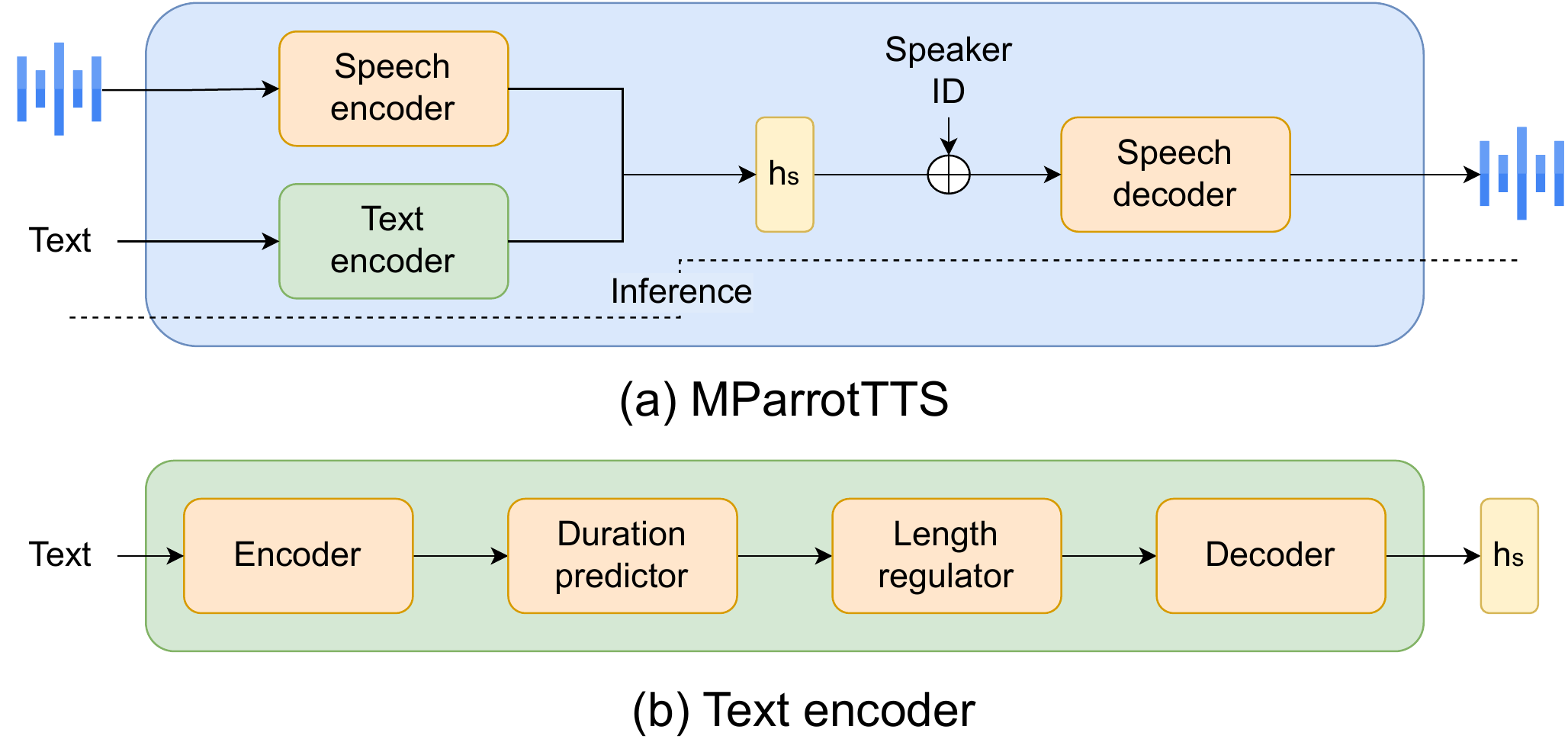}
  \caption{Illustration of \ourmodel~framework. (a) \ourmodel~architecture. The Speech Encoder is a pre-trained network, whereas Text encoder and Speech decoder are trainable networks. The speech decoder is trained from unpaired raw audio. (b) \ourmodel~uses a non-autoregressive text-encoder, shared across all languages.  }
  \label{fig:block diagram}
\end{figure}

Previous work has addressed these challenges in various ways. One of the primary ideas is to learn multilingual TTS by explicitly conditioning the decoder on learnable language and speaker embeddings~\cite{liu2019cross,zhang2019learning}. Speaker embeddings are sometimes obtained using a pre-trained speaker verification network. Additional embeddings capturing other properties like tone have also been utilized~\cite {liu2019cross}. However, the quality remains below par if the text representation is entangled with speaker or the language information, specially when only a few speakers are used for each language. For this reason, several recent methods attempt to disentangle the text representation from language and speech information. Disentanglement is typically attained by back-propagating reverse gradients through an explicit language or speaker classification branch~\cite{zhang2019learning,cho2022sane,nekvinda2020one}. 

Nekvinda and Dušek~\cite{nekvinda2020one} propose Meta-TTS, which uses a contextual parameter generation using language-specific convolutional text encoders. Cho~\etal~\cite{cho2022sane} build upon Meta-TTS by employing an additional speaker regularization loss. They also investigate the effect of using different formats for the textual input (phoneme, character, etc.). Some efforts have explored aspects of knowledge sharing~\cite{prakash2019building} and knowledge distillation~\cite{xu2020lrspeech} for multilingual TTS. More recently, Wu~\etal~\cite{wu2022multilingual} uses a data augmentation technique based on a pre-trained voice conversion model to generate insufficient paired data for a pre-selected set of languages.

Despite the fact that the aforementioned endeavors have played a crucial role in expanding the realm of multilingual TTS, there are still certain limitations that persist. The majority of them~\cite{nekvinda2020one,chen2019cross,zhang2019learning,zhang2020unsupervised} rely on Tacotron~\cite{wang2017tacotron} as their backbone, which is known to suffer from word alignment and word skipping/repetition errors. As pointed out by Nekvinda and Dušek~\cite{nekvinda2020one}, most prior multilingual TTS models limit to 2-3 languages simultaneously or require vast amounts of data to be trained. Furthermore, the training remains challenging in low-resource settings, especially in the absence of paired data from multiple speakers in each language.

In this work, we propose \ourmodel,~which exploits the recent success in self-supervised audio representations~\cite{lee2021textless} to address the aforementioned limitations. \ourmodel architecture is illustrated in Figure~\ref{fig:block diagram}; in the first self-supervised phase of training, a speech-encoder converts audio into discretized sound units, which are then reconstructed back to audio using a speech-decoder. In the second phase, a text encoder maps the input text sequence to a sequence of discrete sound units corresponding to the paired audio. We use HuBERT~\cite{hsu2021hubert,lee2021textless} based embeddings as the self-supervised representation which has proven successful in Speech-to-Speech translation tasks \cite{polyak2021speech,jia-etal-2022-cvss}. A significant advantage of using HuBERT-based embeddings is that they ignore speaker and acoustic information and encode only the speech content \cite{polyak2021speech}. This characteristic is a boon in a low-resource setting, allowing the text-encoder to be trained on a small amount of paired data from a single speaker. Other benefits encompass the ability to handle audio noise during training and simplified training, eliminating the necessity for domain adversarial training \cite{cho2022sane,ganin2016domain}. \ourmodel also employs knowledge sharing by training a unified text-encoder for multiple languages, where we explore character and phoneme-based inputs.

% Robustness against noise in the paired training data is conferred by it and removes requirement on adversarial 

% The model first learns to capture the phonetic variations embedded in a sound by independently training an unsupervised speech decoder module conditioned on speaker and language identity using discrete audio representations for all target speakers. A pre-trained self supervised module is tasked to learn these audio representations. Next, the model learns to generate text embedding sequence using a text encoder module which mimics the ground-truth audio representations.

\ourmodel works in extremely low resource regime~\cite{xu2020lrspeech} and we show that using $5$ hours of single speaker paired data for each language is sufficient for training the unified model, capable of synthesizing high-quality speech corresponding to any unseen text on each of the seen language and the speaker. %\ourmodel adapts to a new speaker simply using a few minutes of unpaired audio samples, irrespective of the spoken language. For instance, 
We show that \ourmodel can generate high-fidelity speech for multiple voices from several European and Indic datasets, whose paired data is not used in training the model in any of the six languages explored in our study. Interestingly, \ourmodel works well even for languages unseen during the self-supervised audio representation learning.

\section{Method}
The proposed \ourmodel is shown in Figure \ref{fig:block diagram}. It consists of three modules, a speech-encoder, a text-encoder, and a speech-decoder. We adopt the speech-encoder and speech-decoder from \cite{polyak2021speech}. The speech-encoder is mHuBERT \cite{lee2021textless}, which maps the input speech signal to discrete sound units, and the speech-decoder is HiFiGAN \cite{kong2020hifi}, which converts the discrete sound units to audio signal. The text-encoder is a Non-autoregressive (\nar) encoder-decoder module that maps the input text representation to the sound units. 

\subsection{Text representation}

Representing text from a set of various languages poses a challenge. 
Traditionally, end-to-end single language TTS models have used either characters \cite{wang2017tacotron} or phonemes \cite{ren2020fastspeech,elias2021parallel} as input text representation. In this paper, we separately evaluate the effects of using these representations to train the text encoder under a multilingual scenario. For a character representation, we extract the tokens using a dictionary created by iterating over the entire text corpus. In contrast, for phoneme representation, we utilize an off-the-shelf phonemizer \cite{bernard2021phonemizer} to extract phonemes belonging to the IPA vocabulary, which are common across languages. A benefit of this step is that the model is language-agnostic, as the model sees the common phonemes across the languages.% Overall, we extract 308 unique character tokens and 632 unique phoneme tokens as a pre-processing step. 

\subsection{Speech-encoder}
We use a pre-trained mHuBERT \cite{lee2021textless} as our speech-encoder $\enc{s}$, which is trained on speech-only data to generate discrete sound units. The model is trained using a Masked Language Model (MLM) loss similar to BERT \cite{devlin2018bert} architecture. mHuBERT is pre-trained on multiple languages showing the model's capability to capture sound units across the languages and generalize to new unseen languages. We also experimented with original HuBERT \cite{hsu2021hubert} embeddings and found that mHuBERT fares better in the studied multilingual scenario. For a given audio signal X, the speech embedding sequence~$\hmb{s}$ are computed with the speech-encoder as:
\begin{align}
  \hmb{s} &= (h_1,....,{h_{\widehat{T}}}) \coloneqq \enc{s} (X),
\end{align}
Where $X = (x_1,x_2,..,x_T)$ is an input speech signal, $\widehat{T} = T/320$ indicates down-sampling of the audio signal. The sound units $h_i \in \{1,\dots,k\}$ where $k = 1000$ is the number of clusters in the mHuBERT's clustering step.

\subsection{Text-encoder}
Text-encoder $\enc{t}$ is a \nar-~based encoder-decoder architecture that maps text representations (character/phoneme tokens) $P=(p_1,p_2,..,p_N)$ to sound unit sequence $\hmb{p}$, predicted as:
\begin{align}
  \hmb{p} &= (h_1,....,{h_{\widehat{P}}}) \coloneqq \enc{t} (P),
\end{align}
where $\widehat{P}$ is the number of sound units. 
The encoder and decoder are feed-forward transformer blocks with self-attention \cite{vaswani2017attention}, and $1$ dimensional convolutions similar to Fastspeech2 \cite{ren2020fastspeech}. The encoder encodes the text representation $P$ into a sequence of fixed dimensional token embeddings. These token embeddings are then passed to the duration predictor and length regulator to determine the output length $\widehat{P}$. The decoder simultaneously predicts all $\widehat{P}$ units of~$\hmb{p}$. We note that our model's performance is not enhanced by conditioning the decoder with an additional language embedding, and as such, we choose to refrain from doing so. The text-encoder$~\enc{t}$ is trained to mimic its output~$\hmb{p}$ to be similar to~$\hmb{s}$, sound units of corresponding audio signal extracted from a pre-trained$~\enc{s}$ speech-encoder. The text-encoder thus requires parallel data (P,X) to learn the mapping from text content to speech content. 

\subsection{Speech-decoder}
We use a modified version of HiFiGAN \cite{kong2020hifi} decoder to decode embeddings to speech. During training, we concatenate~$\hmb{s}$ with a one-hot speaker vector~$\hmb{spk}$ to generate~$\hmb{} = (\hmb{s}, \hmb{spk})$. The speech-decoder has a generator $G$ and a discriminator $D$. The generator $G$ takes~$\hmb{}$ as an input and passes it through a series of transposed convolutions and residual blocks with dilation to generate a synthesized speech signal $\widehat{X}$. The discriminator $D$ has two sets of networks. The Multi-period discriminator handles the periodic portion of input audio, whereas the multi-scale discriminator captures long-term dependencies and patterns. Overall, the discriminator evaluates and distinguishes the synthesized $\widehat{X}$ from the original speech signal $X$.

\section{Experiments}
\subsection{Dataset}
We collate our multilingual dataset using various publicly available corpora containing samples from multiple speakers in six languages: (1) $80.76$ hours of Hindi and Marathi \cite{SYSPIN} from $2$ speakers, respectively; (2) $71.69$ hours of German \cite{MAILabs} from $3$ speakers; (3) $83.01$ hours of Spanish \cite{MAILabs} from $3$ speakers; (4) $10.70$ hours of French \cite{honnet2017siwis} from $1$ speaker; (5) $23.92$ hours of English \cite{ljspeech17} from $1$ speaker. Overall the dataset comprises of $354.12$ hours of paired TTS data from $12$ speakers. We resample all speech samples to $16$~$\khz$. %The speech decoder training uses speech-only data from all $12$ speakers, whereas the text encoder training requires only $30$ hours of paired TTS data (i.e. $5$ hours of single-speaker data for every language).

\subsection{Speech-encoder training}
We use a pretrained $12$-layer transformer-based mHuBERT \cite{lee2021textless} architecture trained for $3$ iterations on $13.5$k hours of English, Spanish and French data from VoxPopuli unlabelled speech corpus \cite{lee2021textless,wang2021voxpopuli}. We extract a $768$ dimensional continuous vector from the eleventh activation layer similar to \cite{lee2021textless}. These vectors are further discretized to~$\hmb{s}$ units using a k-means clustering algorithm with $k=1000$.
% The continuous vector is further discretized to extract~$\hmb{s}$ units using a single codebook trained on 1000 $k$-means cluster. 

\subsection{Text-encoder training}
To demonstrate the efficacy of our model in a low-resource setting, we only use $5$ hours of paired data for a single speaker in each language to train the text encoder (using mere 30 hours of paired data across all languages in total). We report the evaluation metrics for \textit{seen speakers} where the model has seen the speaker paired data and \textit{unseen speakers} whose paired data is not used to train the text-encoder.

To evaluate the performance of the TTS systems on various text representations, we train two variants of the text-encoder, the character text-encoder (CTE) and the phoneme text-encoder (PTE). The first variant uses character tokens across the languages to learn sound units, while the second uses phoneme tokens. Additionally, we employ Deep Forced Aligner \cite{DFA} to align ground-truth speech and input text representations to train the duration predictor. We use cross-entropy loss with $1000$ classes to predict $\hmb{p}$.  
% To validate the effectiveness of UMM-TTS on less paired data, our models are trained only on $5$ hours of single-speaker data for each of the $6$ languages during training. 
% For evaluation purposes, we term the speakers seen to the text-encoder as \textit{seen speakers} and the remaining as \textit{unseen speakers}.

\subsection{Speech-decoder training}
We train a multi-speaker speech-decoder using speech-only data from all $12$ speakers since training this model doesn't require any paired data. First, we extract the sound units~$\hmb{s}$ from the source audio using the pre-trained speech-encoder~$\enc{s}$. During decoder training, we concatenate the sound units~$\hmb{s}$ with one-hot speaker vector~$\hmb{spk}$. However, during inference, we pass the generated sound units~$\hmb{p}$ from the text-encoder along with the target speaker's one-hot vector to synthesize speech. 

\subsection{Baseline models}
% In this paper, we seek to investigate two objectives with the use of two baselines. 
%examine the impact of using self-supervised discrete sound units in multilingual synthesis by comparing  
% To achieve this, we train a multi lingual, multi speaker \baselinefs~(FS2)  on all $12$ speakers in all languages. 

We compare against the publicly available state-of-the-art multilingual TTS approaches. First, we train a fully supervised multi-lingual, multi-speaker \baselinefs~(FS2) model~\cite{chien2021investigating}. We train the \baselinefs~model on the entire dataset (approximately $12$ times more paired data than \ourmodel),~seeing all the $12$ speakers. Only 100 samples for each speaker are kept aside for validation of \baselinefs~model. 

We also compare against the \baselineml~\cite{nekvinda2020one}, a meta learning-based TTS model which uses language-specific contextual parameter generation. We use their released checkpoint and compare the naturalness of the synthesized speech on the shared set of languages. Specifically, we compare their model across the French, Spanish, and German languages. 

Please note that both FS2 and \baselineml~models are fully supervised; they use all the available paired text-speech pairs to train their end-to-end TTS. In contrast, \ourmodel~only uses a small portion of the paired data. 

% We take a pretrained model trained on French, Spanish and German languages and compare against our model's performance in those three languages.

% aim to comprehend the optimal approach for synthesizing a human-like speech output in a specific language using an end-to-end multilingual TTS system. To accomplish this, we compare against \baselineml-based TTS model [9] and conduct evaluations on the common languages trained across \baselineml~model and our model.

\subsection{Evaluation metrics}
% We evaluate the model's performance in synthesizing multi lingual speech using naturalness, speaker similarity and 

We use the Mean Opinion Score (MOS) to evaluate the quality of synthesized speech, where the participants are asked to rate the samples based on their naturalness. The user is asked to rate the naturalness of the audio where $5$ is ``completely natural,'' and $1$ is ``completely unnatural''. And to evaluate the model's ability to capture speaker information, we use speaker similarity MOS, where the participants rate the similarity of a synthesized sample with a ground truth sample from the same speaker.
We also assess the model's performance in cross-lingual synthesis, where a speaker's voice in one language is used to synthesize content in a different language. In addition to evaluating the naturalness of the synthesized voice, participants were also requested to identify the native accent of the speaker. This experiment was conducted to assess the effectiveness of our proposed setup in preserving the nuances of the native speaker when speaking foreign content. 

We conduct our evaluations using the Absolute category rating scale, in which the subjects are asked to provide ratings on a scale of $1$ to $5$, with a $1$ point increment. We collect evaluations from five native speakers for each language across all the subjective listening tests. We randomly select five sentences from the test set, and each participant is asked to rate the samples synthesized using our model and the baseline model along with the ground truth audio. For English and French languages, we only use single speaker data; hence we exclude these two languages as part of \textit{unseen speaker} evaluation. 

\section{Results and Discussion}

%\subsection{Speech synthesis quality}

\begin{table}[t]
  \caption{Comparison of Naturalness MOS on seen speakers with FS2 model}
  \label{tab:mos-naturalness-seen}
  \centering
  \resizebox{0.47\textwidth}{!}{
  \begin{tabular}{rrrrrrr}
    
    \toprule
    \multicolumn{1}{c}{\textbf{Language}} & 
    \multicolumn{1}{c}{\textbf{GT}} &
    \multicolumn{1}{c}{\textbf{CTE}} &
    \multicolumn{1}{c}{\textbf{PTE}} &
    \multicolumn{1}{c}{\textbf{FS2}}\\
    \midrule
    Hindi & $3.78$ $\pm$ $0.14$ & $\textbf{3.33}$ $\pm$ $\textbf{0.19}$ & $3.22$ $\pm$ $0.15$ & $3.33$ $\pm$ $0.12$ \\
    Marathi & $4.81$ $\pm$ $0.07$ & $\textbf{3.78}$ $\pm$ $\textbf{0.12}$ & $3.04$ $\pm$ $0.19$ & $3.59$ $\pm$ $0.15$ \\
    German & $3.54$ $\pm$ $0.20$ & $3.33$ $\pm$ $0.19$ & $\textbf{3.58}$ $\pm$ $\textbf{0.12}$ & $3.21$ $\pm$ $0.16$ \\
    French & $3.83$ $\pm$ $0.19$ & $2.23$ $\pm$ $0.14$ & $\textbf{4.17}$ $\pm$ $\textbf{0.19}$ & $3.50$ $\pm$ $0.16$ \\
    English & $4.20$ $\pm$ $0.12$ & $3.11$ $\pm$ $0.11$ & $\textbf{3.50}$ $\pm$ $\textbf{0.10}$ & $2.50$ $\pm$ $0.18$ \\
    Spanish & $3.67$ $\pm$ $0.12$ & $3.5$ $\pm$ $0.21$ & $\textbf{3.67}$ $\pm$ $\textbf{0.20}$ & $2.50$ $\pm$ $0.21$ \\

    \bottomrule
  \end{tabular}
}
\end{table}

\begin{table}[t]
  \caption{Comparison of Naturalness MOS on unseen speakers with FS2 model}
  \label{tab:mos-naturalness-unseen}
  \centering
  \resizebox{0.47\textwidth}{!}{
  \begin{tabular}{rrrrrrr}
    \toprule
    \multicolumn{1}{c}{\textbf{Language}} & 
    \multicolumn{1}{c}{\textbf{GT}} &
    \multicolumn{1}{c}{\textbf{CTE}} &
    \multicolumn{1}{c}{\textbf{PTE}} &
    \multicolumn{1}{c}{\textbf{FS2}}\\
    \midrule
    Hindi & $4.22$ $\pm$ $0.18$ & $\textbf{3.28}$ $\pm$ $\textbf{0.19}$ & $3.05$ $\pm$ $0.20$ & $3.22$ $\pm$ $0.21$ \\
    Marathi & $4.48$ $\pm$ $0.13$ & $\textbf{3.63}$ $\pm$ $\textbf{0.18}$ & $3.11$ $\pm$ $0.18$ & $3.15$ $\pm$ $0.19$ \\
    German & $3.17$ $\pm$ $0.22$ & $2.72$ $\pm$ $0.23$ & $\textbf{3.55}$ $\pm$ $\textbf{0.20}$ & $2.05$ $\pm$ $0.22$ \\
    Spanish & $3.67$ $\pm$ $0.19$ & $3.17$ $\pm$ $0.17$ & $\textbf{3.33}$ $\pm$ $\textbf{0.18}$ & $3.17$ $\pm$ $0.19$ \\

    \bottomrule
  \end{tabular}
  }
\end{table}

\textbf{Naturalness MOS} %As shown in Table~\ref{tab:mos-naturalness-seen} and Table~\ref{tab:mos-naturalness-unseen}, our model performs much better than the FS2 baseline with huge margin, even though it has seen just $1/6$th of the paired data used for FS2 model training.
Based on the results presented in Table~\ref{tab:mos-naturalness-seen} and Table~\ref{tab:mos-naturalness-unseen}, our model demonstrates significantly superior performance compared to a multilingual FS2 model. Notably, our model's success is even more impressive considering that it was trained on only a fraction, specifically $1/12$th, of the paired data used to train the FS2 model. %Especially this shows the efficacy of ~\ourmodel in a low-resource scenario where collecting paired data is difficult and costly. 
This indicates the viability of \ourmodel~in a low-resource scenario where collecting paired data is difficult and costly. We also observe that the naturalness in different languages varies with the input text representation. For Indic languages (Hindi and Marathi), representing the text in characters (CTE) performs better than phonemes (PTE). Whereas for English and the European Languages (German, French, and Spanish), phoneme representation PTE serves better. With the appropriate choice of tokenization scheme (CTE or PTE), our model consistently outperforms FS2 on both seen and unseen speakers, even though no paired data is used for unseen speakers. Another interesting observation is that for some of the European languages~\ourmodel~fares better than the ground truth. We believe the noise in the ground-truth data hampers its naturalness score. Since mHubert embeddings ignore/suppress the ambient information, it imparts noise robustness to~\ourmodel,~improving the naturalness over the ground-truth audio in noisy scenarios.

% We report the naturalness in Table 
% Among the multi-speaker languages, we randomly select German's \textit{Angela merkel} and Spain's \textit{Tux} as \textit{seen speakers}, and the rest as \textit{unseen speakers} for training text encoder. Table \ref{tab:mos-naturalness-speakersimilarity} provides empirical evidence that our setup significantly preserves target speaker characteristics as compared to \baselinefs~that relies on mel-based vocoder to synthesize speech. 

% Table \ref{tab:mos-naturalness-seen} and Table \ref{tab:mos-naturalness-unseen} shows the naturalness MOS on \textit{seen speakers} and \textit{unseen speakers}. Our findings indicate that the character tokenization should be the preferred text representation for Indic languages, while for European languages, tokenizing phonemes yields better results. By utilizing the preferred tokenization scheme, we consistently observe that our model performs as well as or better than the \baselinefs~even for the unseen speakers whose paired data was not used during training.

%We also report the MOS of UMM-TTS with \baselineml model on the common language set (French, Spanish and German) used in both the models in Table \ref{tab:mos-naturalness-meta}. 
Table \ref{tab:mos-naturalness-meta} includes our findings on the naturalness of \ourmodel~as assessed by MOS scores, when compared against the \baselineml~model on the shared set of languages (French, Spanish, and German). While our model gives comparable results for German, it significantly outperforms \baselineml~in French and Spanish. The MOS scores for German remain low for~\ourmodel~because of low-quality training data (even ground truth MOS scores are low in Table~\ref{tab:mos-naturalness-seen} and Table~\ref{tab:mos-naturalness-unseen}). \baselineml~uses 20 hours of paired data from 40 German speakers for their training. They also perform extensive pre-processing, which includes careful data selection methods. As our model is intended to operate in a low-resource setting where gathering paired data poses a challenge, we refrain from applying any data-cleaning techniques and limit the text-encoder training to five hours of noisy data from a single speaker.

%We note here that our ground-truth German data has background noise, as also corroborated by the participants in Table \ref{tab:mos-naturalness-seen} and Table \ref{tab:mos-naturalness-unseen} making it sound un-natural. 

% We believe the performance gain observed with the \baselineml~model for German can be attributed to its pre-processing step, which includes careful data selection methods. As our model is intended to operate in a low-resource setting where gathering paired data poses challenges, we refrain from applying any data cleaning techniques. 

\begin{table}[t]
  \caption{Comparison of Naturalness MOS with \baselineml~model}
  \label{tab:mos-naturalness-meta}
  \centering
  \begin{tabular}{rrrrrrr}
    \toprule
    \multicolumn{1}{c}{\textbf{Language}} & 
    \multicolumn{1}{c}{\textbf{Our model}} &
    \multicolumn{1}{c}{\textbf{\baselineml~\cite{nekvinda2020one}}} \\
    \midrule
    French & $\textbf{4.08}$ $\pm$ $\textbf{0.19}$ & $2.58$ $\pm$ $0.18$ \\
    Spanish & $\textbf{3.75}$ $\pm$ $\textbf{0.16}$ & $3.00$ $\pm$ $0.20$ \\
    German & $3.33$ $\pm$ $0.19$ & $\textbf{3.50}$ $\pm$ $\textbf{0.18}$ \\
    \bottomrule
  \end{tabular}
  
\end{table}

\textbf{Speaker Similarity} We conducted a comparative analysis of our model's ability to capture speaker identity as opposed to that of the FS2 model. Speaker similarity is reported on a scale of $1$ to $5$, where $1$ corresponds to ``Not at all similar,'' and $5$ corresponds to ``extremely similar''. Our results, presented in Table \ref{tab:mos-naturalness-speakersimilarity}, consistently demonstrate the superiority of our model over FS2, indicating its effectiveness in separating speaker and content information. This is attributed to the decoder being conditioned solely on speaker ID and the shared sound unit space, which is common across all languages.

% Our model 
% We attribute this to be because the 

\begin{table}[t]
  \caption{Comparison of speaker similarity MOS with FS2 model}
  \label{tab:mos-naturalness-speakersimilarity}
  \centering
  \begin{tabular}{rrrrrrr}
    \toprule
    \multicolumn{1}{c}{\textbf{Language}} & 
    \multicolumn{1}{c}{\textbf{Our model}} &
    \multicolumn{1}{c}{\textbf{FS2}}\\
    \midrule
    Hindi & \textbf{$4.29$ $\pm$ $0.18$} & $3.92$ $\pm$ $0.21$ \\
    Marathi & \textbf{$4.21$ $\pm$ $0.16$} & $3.83$ $\pm$ $0.08$ \\
    German & \textbf{$4.09$ $\pm$ $0.11$} & $3.25$ $\pm$ $0.14$ \\
    French & \textbf{$3.87$ $\pm$ $0.20$} & $3.50$ $\pm$ $0.19$ \\
    English & \textbf{$3.94$ $\pm$ $0.18$} & $3.00$ $\pm$ $0.19$ \\
    Spanish & \textbf{$4.33$ $\pm$ $0.17$} & $3.50$ $\pm$ $0.19$ \\

    \bottomrule
  \end{tabular}
  
\end{table}

\textbf{Cross lingual synthesis}
We also assess the model's performance in synthesizing the samples of a speaker in a different language. Table \ref{tab:mos-naturalness-crosslingual} presents the results of a study that compares the naturalness of MOS in a cross-lingual setting. The first column shows the speaker-text pair, where the speaker of a native language attempts to speak a different non-native language. We report the MOS scores for a few speaker-text pairs. 

Our model exhibits significantly higher ratings than FS2, demonstrating the flexibility of our model to generate sound units from any given text and synthesize speech for any target speaker of a different native language. In addition, more than $90$ \% of the participants were able to discern the nativity of the synthesized speech produced by both the models.

% The first column shows a native speaker who tries to read a text in a non-native language, with the speaker-text pairs selected randomly. 

\begin{table}[t]
  \caption{Comparison of Naturalness MOS for cross-lingual speech synthesis}
  \label{tab:mos-naturalness-crosslingual}
  \centering
  \begin{tabular}{rrrrrrr}
    \toprule
    \multicolumn{1}{c}{\textbf{Speaker-Text}} & 
    \multicolumn{1}{c}{\textbf{Our model}} &
    \multicolumn{1}{c}{\textbf{FS2}}\\
    \midrule
    Hindi-Spanish & $3.87$ $\pm$ $0.22$ & $3.25$ $\pm$ $0.19$ \\
    Marathi-English & $3.63$ $\pm$ $0.21$ & $3.5$ $\pm$ $0.22$ \\
    French-Hindi & $4.07$ $\pm$ $0.12$ & $2.71$ $\pm$ $0.21$ \\
    Spanish-German & $4.14$ $\pm$ $0.20$ & $2.29$ $\pm$ $0.21$ \\
    English-German & $3.57$ $\pm$ $0.15$ & $2.43$ $\pm$ $0.18$ \\
    English-Hindi & $3.57$ $\pm$ $0.19$ & $2.57$ $\pm$ $0.18$ \\
    French-German & $3.93$ $\pm$ $0.17$ & $2.71$ $\pm$ $0.18$ \\
    Spanish-French & $3.71$ $\pm$ $0.18$ & $2.57$ $\pm$ $0.17$ \\
    Hindi-Marathi & $4.13$ $\pm$ $0.21$ & $3.25$ $\pm$ $0.19$ \\
    Marathi-French & $2.87$ $\pm$ $0.19$ & $2.75$ $\pm$ $0.18$ \\

    \bottomrule
  \end{tabular}
  
\end{table}

Collectively, our findings indicate that \ourmodel~possesses the ability to produce high-quality, multilingual, and multi-speaker speech that effectively captures the distinctive characteristics of individual speakers, even when synthesized for a different language.

\section{Conclusion}
In this paper, we present \ourmodel,~a unified multilingual self-supervised learning-based text-to-speech (TTS) synthesis framework that has the significant potential to enhance the TTS capability on low-resource languages. We demonstrate that high-fidelity speech can be obtained by utilizing a minimal amount of single-speaker training data per language. Our approach enables generalization to various voices within a particular language without requiring paired data for each voice. The experimental findings have shown that the quality of speech synthesis is significantly influenced by the text representations used, such as characters or phonemes, particularly in diverse linguistic contexts. Our experimental outcomes have revealed that our model outperformed the baseline \baselinefs~and \baselineml~models, in terms of naturalness and speaker similarity. In addition, our method has effectively showcased the ability for cross-language synthesis (to produce speech in a language that is not native) while maintaining the speaker-specific nuances and delivery style.

One limitation of our approach is that the duration predictor in the non-autoregressive model might lean towards the style of the single seen speaker. Future research will investigate methods to obtain speaker adaptive duration prediction or allow levers to control the duration prediction. Our current self-supervised backbone is trained on limited set of languages. A direction of future exploration would be to fine-tune self-supervised HuBert-based embeddings using the raw audio from a diverse set of languages (South Asian, Latin, English, etc.) to have a more comprehensive quantized code set of sound units. Another area of future work will be releasing an open-source, multilingual TTS model, allowing the community to apply the investigated ideas to other resource-scarce, less studied, and less privileged languages. 

% Moreover, our approach has successfully demonstrated the capacity to generate speech in a non-native language, while preserving the desired subtlety and speaking style.

% exclusively audio-based information to enhance the learning of discrete sound units in our suggested text-encoder, leading to a further enhancement in the quality of synthetic speech. 

\bibliographystyle{IEEEtran}
\bibliography{mybib}

\end{document}